# The Scientific Legacy of Apollo

**Ian A. Crawford**, Department of Earth and Planetary Sciences, Birkbeck College, University of London (i.crawford@bbk.ac.uk).

**Abstract**
On the 40$^{th}$ anniversary of the last human expedition to the Moon, I review the scientific legacy of the Apollo programme and argue that science would benefit from a human return to the Moon.

**Introduction**

This December marks 40 years since the last human beings to set foot on the Moon, Gene Cernan and Harrison "Jack" Schmitt of Apollo 17, left the lunar surface and returned safely to Earth. This anniversary alone would have justified a retrospective look at the legacy of the Apollo project, but it has been given additional poignancy by the death earlier this year of Neil Armstrong, the first man to set foot on the lunar surface with Apollo 11 in July 1969. The history of the Apollo project, and its geopolitical motivation within the context of the Cold War, is of course well documented (e.g. Chaiken 1994; Burrows 1998; Orloff and Harland 2006) and need not be repeated here. However, although the *scientific* legacy of Apollo has also been well-documented (e.g. Heiken et al. 1991; Wilhelms 1993; Beattie 2001), and is generally well-known within the lunar science community, I have found that it is still underappreciated by many astronomers, and even some planetary scientists who are not directly involved in lunar studies. That, at any rate, is my justification for taking this opportunity to give a brief review of Apollo science.

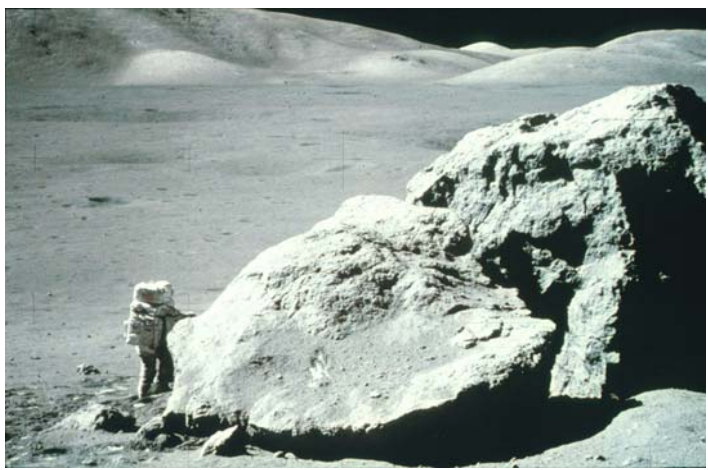

Fig. 1. One of the last two men on the Moon: Harrison Schmitt stands next to a large boulder at the Apollo 17 Station 6 locality in December 1972. Note the sampling of regolith on the boulder's upper surface (NASA).

In the three and a half years between Armstrong's 'first small step' in 1969 and the departure of Cernan and Schmitt from the Taurus-Littrow Valley (Fig. 1) in 1972, a total of twelve astronauts explored the lunar surface in the immediate vicinity of six Apollo landing sites (Fig. 2). The total cumulative time spent on the lunar surface was

12.5 days, with just 3.4 days spent performing extravehicular activities (EVAs) outside the lunar modules (Orloff and Harland 2006). Yet during this all-too-brief a time samples were collected, measurements made, and instruments deployed which have revolutionised lunar and planetary science and which continue to have a major scientific impact today.

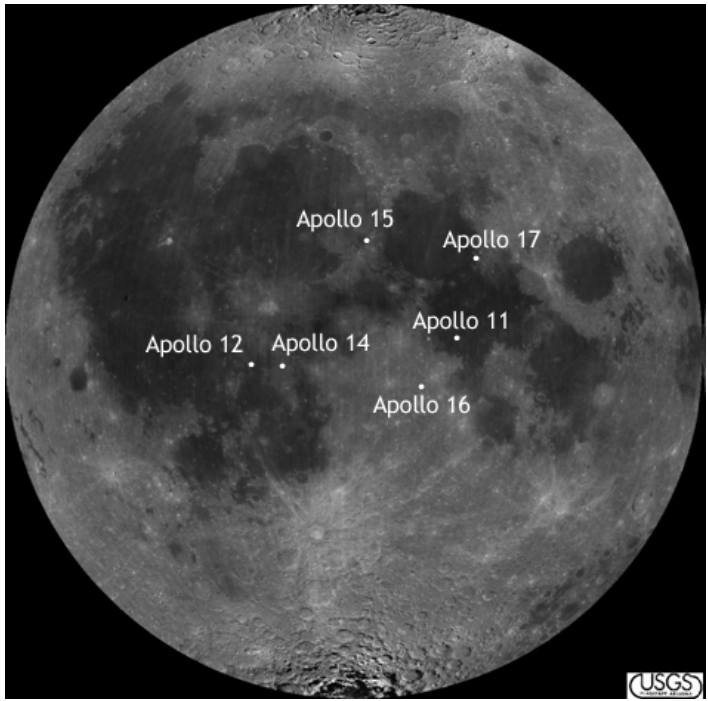

Fig 2. The Apollo landing sites. Note their restriction to the central part of the nearside – there is a lot more of the Moon to explore! (USGS/Dr K.H. Joy).

**Exploration Efficiency**

In their cumulative 12.5 days (25 man-days) on the lunar surface, the twelve Apollo moonwalkers traversed a total distance of 95.5 km from their landing sites (heavily weighted to the last three missions that were equipped with the Lunar Roving Vehicle), collected and returned to Earth 382 kg of rock and soil samples (from over 2000 discrete sample localities), drilled three geological sample cores to depths greater than 2 m (plus another five 2-3 m cores for the heat-flow experiments), obtained over 6000 surface images, and deployed over 2100 kg of scientific equipment. These surface experiments were supplemented by wide-ranging remote-sensing observations conducted from the orbiting Command/Service Modules, which are of course equally part of the Apollo legacy. Interested readers will find comprehensive summaries of all the Apollo experiments given by Wilhelms (1993), Beattie (2001), and Orloff and Harland (2006).

Before moving on to discuss the main scientific results from all this activity, I think it is worth pausing to reflect on the sheer efficiency of the Apollo astronauts as scientific explorers. This may only be immediately obvious to colleagues who themselves have experience of geological fieldwork, and I am happy to provide a personal example. In June 2011, as part of an astrobiology project to assess the potential of the Kverkfjoll sub-glacial volcano in central Iceland as a Mars analogue site (see Cousins and Crawford 2011), myself and five colleagues spent 7 days operating out of a small mountain hut on the Vatnajökull glacier not a whole lot bigger than an Apollo lunar module. During this time we traversed a total distance of approximately 10 km (by foot), made a detailed map our field locality, collected and returned about 25 kg of geological samples, deployed and/or employed about 20 kg of scientific equipment (including a field spectrometer and equipment to make in situ environmental and geochemical analyses of various kinds), and took about 900 images (easier with today's digital cameras than with the bulky Apollo Hasselblads of course); we did not obtain any drill cores or make any geophysical measurements, but then our particular project didn't require these.

I do not think that we were inefficient, and we were in fact well-pleased with what we accomplished (which will result in several peer-reviewed publications), but clearly what we achieved in 42 man-days at one site in Iceland pales into insignificance to what the Apollo astronauts achieved in 25 man-days at six sites on the Moon under far more difficult operating conditions. Based on my own experience I find the field efficiency of the Apollo astronauts to be simply staggering, and I invite other colleagues familiar with field science to compare the efficiency of Apollo with field activities with which they may be familiar. Looking forward, the efficiency demonstrated by the Apollo astronauts augurs well for the scientific returns which may be anticipated from future human expeditions to the Moon and Mars, an argument developed in more detail elsewhere (Crawford 2012).

**Sample analysis**

There can be little doubt that the greatest scientific legacy of Apollo has resulted from analysis of the 382 kg of rock and soil samples returned to Earth (Fig. 3). However, the extent to which the Apollo samples are still central to lunar and planetary science investigations is perhaps one of the most underappreciated aspects of the Apollo legacy. Every year NASA's Curation and Analysis Planning Team for Extraterrestrial Materials (CAPTEM) allocates several hundred samples of Apollo material to investigators in the United States and around the world. In the UK, several groups (notably those at Oxford, Manchester, the Open University, and the author's own group at Birkbeck College), are currently working on Apollo samples for a range of scientific studies.

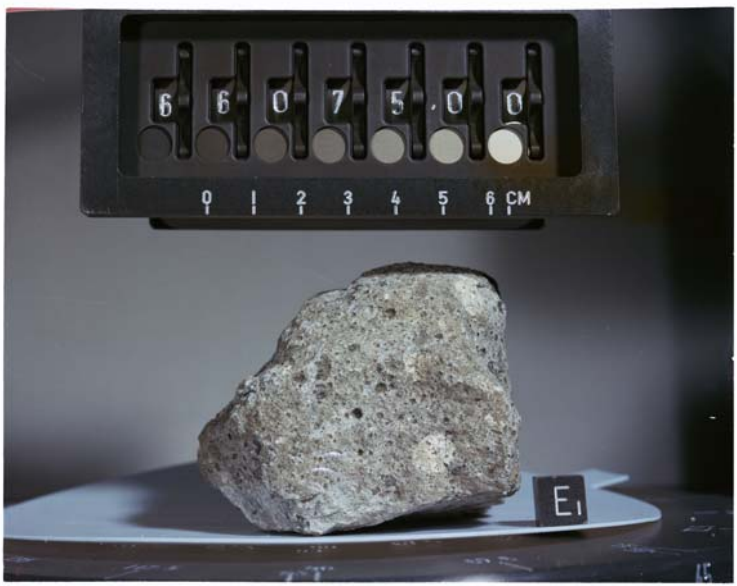

Fig. 3. Apollo 16 sample 66075, a piece of regolith breccia studied by Joy et al. (2012) who identified within it a fragment of a meteorite that struck the lunar surface billions of years ago (NASA).

Probably the most important result based on the Apollo material has been the calibration of the lunar cratering rate, especially over the period 3.2 to 3.8 billion years ago covered by the Apollo samples (reviewed by Stöffler et al. 2006). Only by comparing the density of impact craters on surfaces whose ages have been obtained independently by laboratory radiometric analyses of returned samples is it possible to obtain a calibration of the cratering rate. Analysis of the Apollo samples (supplemented by those obtained by the Soviet Union's Luna robotic missions) has enabled this to be done for the Moon (Fig. 4), which remains the only planetary body for which such a calibration exists. Not only has this facilitated the dating of lunar surfaces from which samples have yet to be obtained, but it is used, with assumptions, to estimate the ages of cratered surfaces throughout the Solar System from Mercury to the moons of the outer planets. In particular, until such time as samples are returned from Mars (an important, but apparently ever-receding, scientific goal of future exploration), extrapolations of the Apollo calibration of the lunar cratering rate remains the only way of dating key events in the history of that planet, including those related to past habitability (Kallenbach et al. 2001). Arguably, this alone would justify the Apollo missions from a scientific point of view.

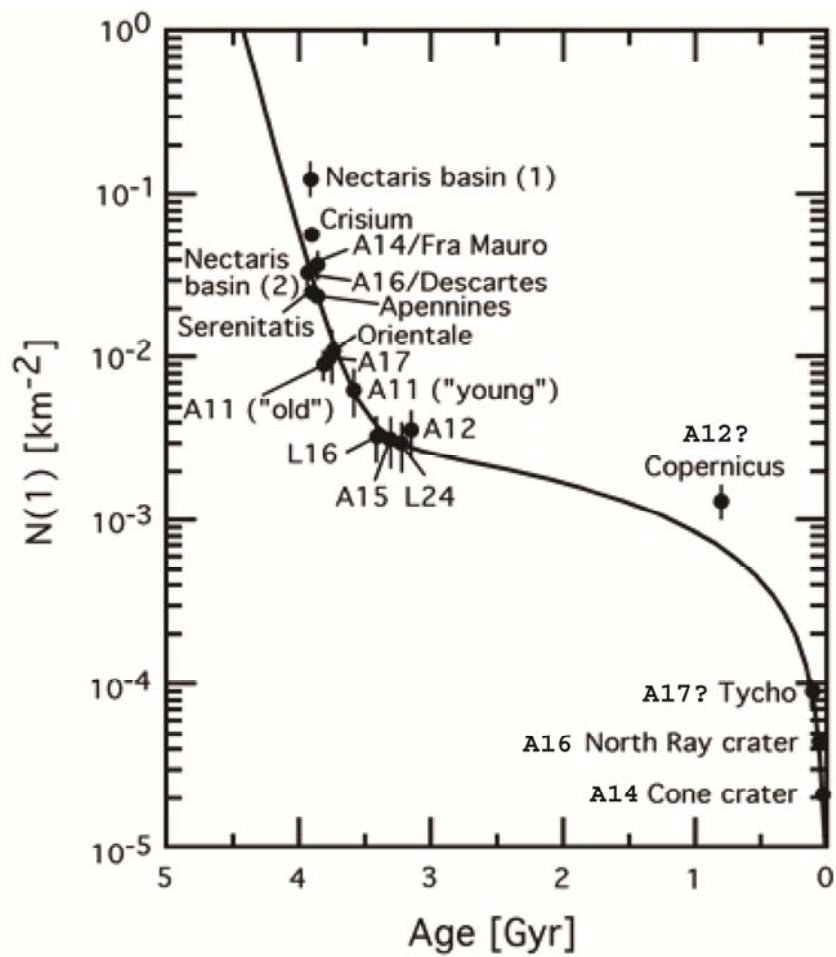

Fig 4. The lunar crater density (number of craters larger than 1 km in diameter per square km) as a function of surface age as calibrated by Apollo (A) and Luna (L) samples (modified from Stöffler et al., 2006; reproduced with permission of the Mineralogical Society of America).

There is however much more that the Apollo samples have revealed about the history of the Moon and the inner Solar System. Perhaps the next most important result of Apollo sample analysis from a planetary science point of view has been the evidence provided for the origin of the Moon. In particular, the discovery that lunar materials have compositions broadly similar to those of Earth's mantle (including nearly identical isotope ratios), but that the Moon is highly depleted in volatiles compared to the Earth and has only a small iron core (a conclusion itself supported by the Apollo geophysics measurements described below), led to the current paradigm that the Moon formed from debris resulting from a giant impact of a Mars-sized planetesimal with the early Earth (e.g. Hartmann and Davis 1975; Jones and Palme 2000; Canup 2004). It is important to realise that constraining theories of lunar origins is of much wider significance for planetary science than 'merely' understanding the origin and early evolution of the Earth-Moon system, important though that is, because it also informs our understanding of the general process of planet formation through the merger of planetesimals in the early Solar System (e.g. Wetherill 1990). It is very doubtful that we would have sufficient geochemical evidence usefully to constrain theories of lunar origins without the quantity and diversity of samples provided by Apollo, and indeed these samples are still being actively exploited for this purpose (e.g. Pahlevan et al 2011; Armytage et al. 2012).

Beyond this, the Apollo samples have been vital to our understanding of the Moon's own geological history and evolution (for recent reviews see Shearer et al 2006; Neal 2009; Jaumann et al 2012). While lunar geology may at first sight appear to be a relatively parochial area of planetary science, it is important to realise that, because its own internal activity largely ceased so long ago, the Moon's surface and interior retain, as if frozen in time, records of planetary differentiation and post-differentiation processes which will have occurred in the early histories of all terrestrial planets. These include records of such key planetary processes as core formation, magma ocean evolution, and primary and secondary crust formation through early magmatic and volcanic activity. In all these respects the Moon acts as a keystone for understanding terrestrial planet evolution more generally (e.g. Head 2012; Kring 2012), and the Apollo samples continue to be used to elucidate important geological processes of relevance both to the Moon itself and wider terrestrial planet evolution (e.g. Borg et al. 2011; Elardo et al. 2011; Shea et al. 2012).

In addition, Apollo samples of the lunar regolith, and regolith breccias formed from it (Fig. 3), have demonstrated the importance of the lunar surface layers as an archive of material which has impacted the Moon throughout its history. These include records of solar wind particles, the cosmogenic products of cosmic ray impacts, and meteoritic debris (see reviews by McKay et al. 1991; Lucey et al. 2006; Crawford et al. 2010). Extracting meteoritic records from lunar regolith samples is especially important for planetary science as it potentially provides a means of determining how the flux and composition of asteroidal material in the inner Solar System has evolved with time (e.g. Joy et al., 2012, and references therein).

Last, but by no means least, the Apollo samples have been used to calibrate remote sensing investigations of the lunar surface. The visible, infrared, X-ray and gamma-ray spectral mapping instruments carried by a host of recent orbital missions to the Moon (notably on the Clementine, Lunar Prospector, Kaguya, Chandrayaan-1 and Lunar Reconnaissance Orbiter spacecraft) have produced a wealth of information regarding the chemical and mineralogical nature of the lunar surface (e.g. Lucey et al. 2000; Jolliff et al. 2000; Pieters et al. 2009; Yamamoto et al. 2010; Glotch et al 2010; Weider et al. 2012). However, although these orbital missions post-date Apollo, and extend compositional measurements to regions of the lunar surface that Apollo did not reach, the reliability of their results largely depends on their calibration against known compositions at the Apollo landing sites. Quite simply, without the 'ground truth' provided by the Apollo samples, it would be difficult to have as much confidence in the results of these remote sensing measurements as we do.

**Geophysics**

Important though the study of the Apollo samples has been, and continues to be, for lunar and planetary science, many other areas of scientific investigation were also performed by the Apollo missions (Beattie 2001). Probably the next most influential set of Apollo experiments were those related to various geophysical investigations, including both passive and active (Fig. 5) seismology studies, surface gravimetry and magnetometry, heat-flow measurements (Fig. 6), and the deployment of laser reflectors to measure the changing Earth-Moon distance and the Moon's physical librations. With the exception of an ineffective seismic experiment sent to Mars on the Viking landers in 1976, the Moon remains the only planetary body apart from Earth on which these geophysical techniques have been applied in situ at the surface.

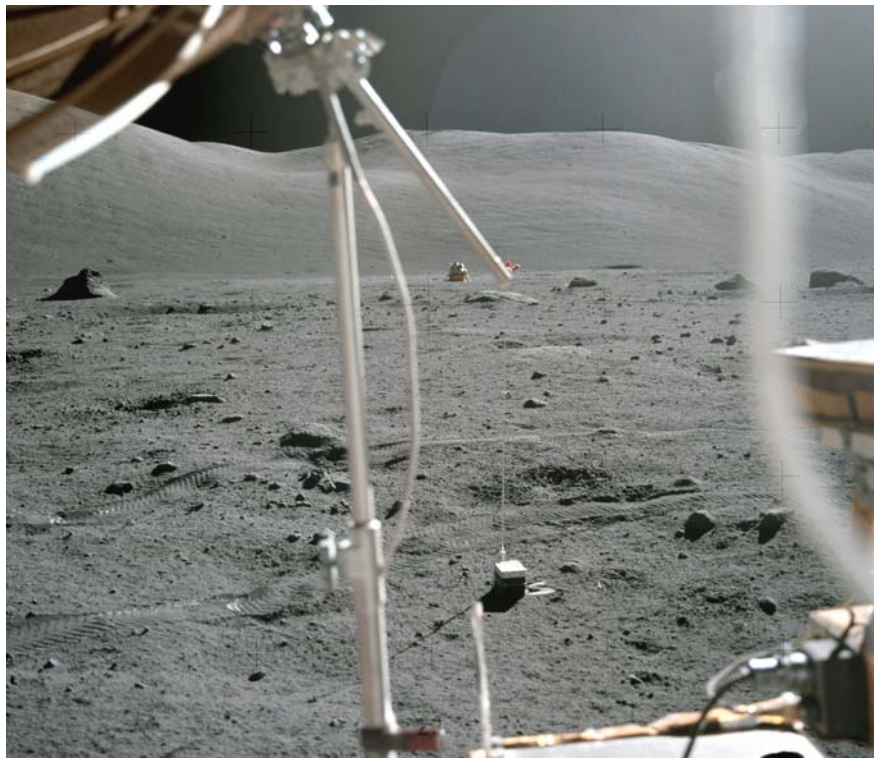

Fig 5. One of eight explosive packages deployed by the Apollo 17 astronauts to provide data for the lunar seismic profiling experiment which measured the thickness of regolith and the underlying lava in the Taurus-Littrow Valley. The Apollo 17 LRV is in the foreground and the lunar module, where a geophone array was deployed to collect the signals, in the middle distance about 300 m away (NASA).

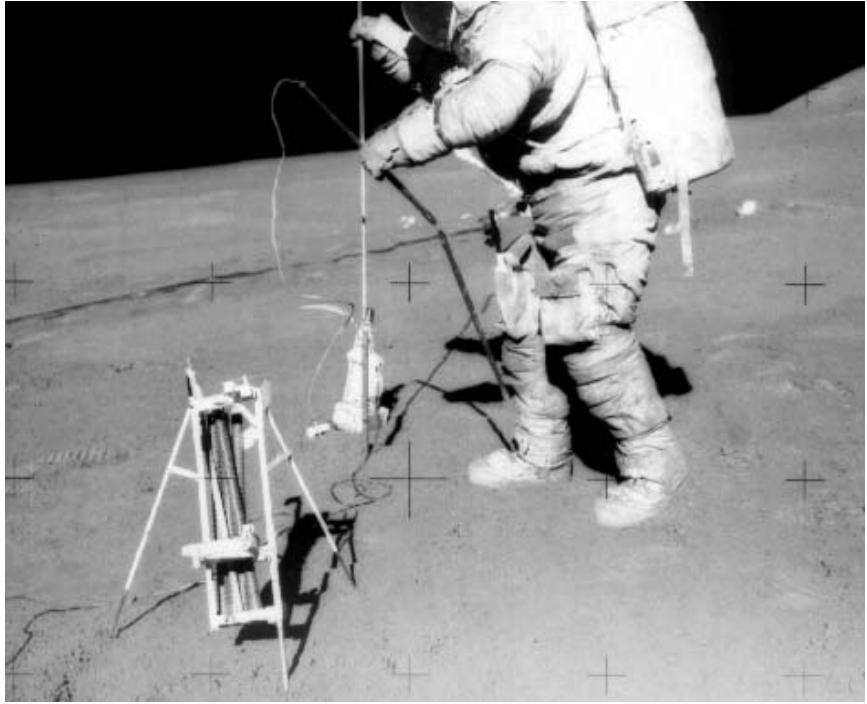

Fig 6. David Scott deploys one of the Apollo 15 heat-flow probes (NASA).

The key results of the Apollo geophysics experiments have been reviewed by Wieczorek et al. (2006) and Jaumann et al. (2012). They include the discovery of natural moonquakes and their exploitation to probe the structure of the nearside crust and mantle, geophysical constraints on the existence and physical state of the lunar core (from both seismic data and laser reflection studies of lunar rotation), the use of active seismic profiling to determine the near-surface structure (Fig. 5), and measurements of the lunar heat-flow at the Apollo 15 and 17 localities. It is important to recognise that, although these data are for the most part over thirty years old (the Apollo seismometers were switched off in 1978), advances in interpretation, and especially in numerical computational techniques, means that they continue to give new insights into the interior structure of the Moon. For example, only last year an apparently definitive seismic detection of the Moon's core, and strong evidence that, like the Earth's, it consists of solid inner and liquid outer layers, was made by a re-examination of Apollo seismic data (Weber et al. 2011).

The deployment of this ambitious range of massive and bulky geophysical instrumentation (as also the large sample return capacity of Apollo) was a beneficiary of, and would arguably have been impossible without, the relatively generous mass budgets that are an inherent feature of human space missions compared to robotic ones (see discussion by Crawford 2012). It therefore seems most unlikely that, without Apollo, our geophysical knowledge of the Moon, and therefore our understanding of the interior structures of small rocky planets more generally, would be anything like as developed as it now is.

**Time to go back**

Looking over the above, I think one could reasonably make the case that Apollo laid the foundations for modern planetary science, certainly as it relates to the origin and evolution of the terrestrial planets. Arguably, the calibration of the lunar cratering rate, and its subsequent extrapolation to estimating surface ages throughout the Solar System, could alone justify this assertion. If one also considers the improvements to our knowledge of lunar origins (and thus the processes involved in forming terrestrial planets), lunar geological evolution (and thus the more general processes of planetary differentiation, core formation, magma ocean crystallization, and crust formation), and the records of solar wind, cosmic rays and meteoritic debris extracted from lunar soils, it is clear that our knowledge of the Solar System would be greatly impoverished had the Apollo missions not taken place. Indeed, the growth of Apollo-derived knowledge is graphically illustrated by the continuing growth of refereed publications based upon it (Fig. 7). At the very least, all this should give pause for thought to those who may still be tempted to agree with the then-Astronomer Royal's comment, voicedon the eve of the Apollo 11 landing, that "from the point of view of astronomical discovery it [the Moon landing] is not only bilge but a waste of money" (Woolley 1969).

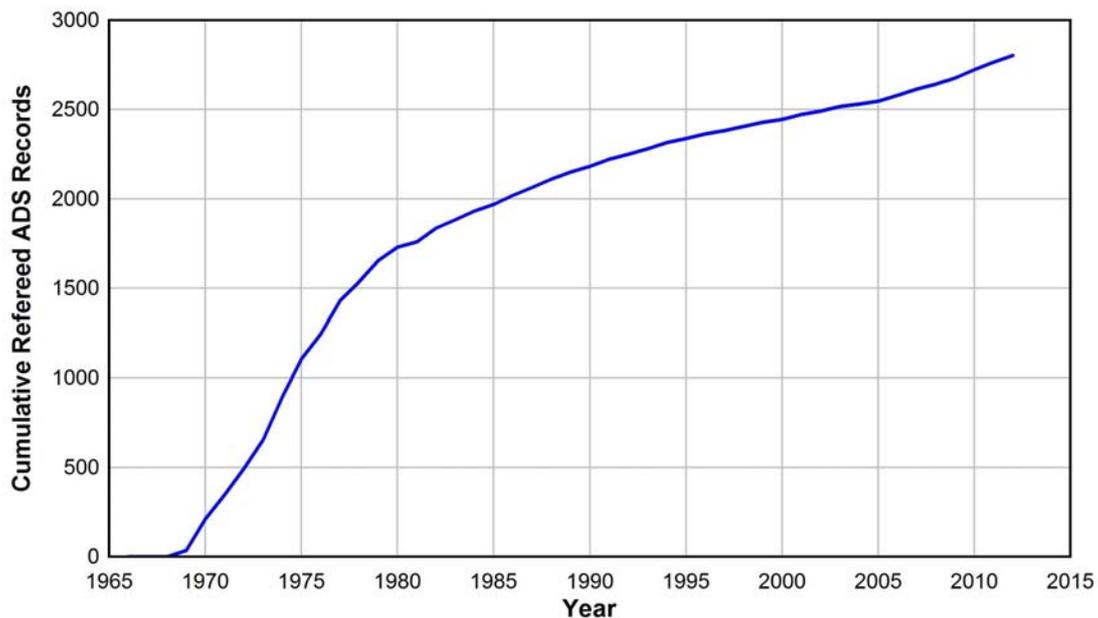

Fig 7. Cumulative number of refereed publications in the ADS database that make use of Apollo data (note that this actually underestimates Apollo-based publications as many are in geological and other journals not covered by the ADS). Recall that this considerable scientific legacy is based on only 25 man-days total contact with the lunar surface (and only 6.8 man-days actually performing EVA activities; see Crawford 2012 for details).

However, despite its rich scientific legacy, it would be a mistake to claim that Apollo did anything more than scratch the surface, both literally and figuratively, of the lunar geological record. With only six landing sites, all at low latitudes on the nearside (Fig. 2), it is clear that much remains to be explored. Moreover, precisely because we have the Apollo legacy as a foundation on which to build, supplemented by recent orbital remote-sensing missions, it is now possible to formulate much more sophisticated lunar exploration strategies than was possible forty years ago. There are now key, specific, scientific questions which can only be addressed by once again returning to the lunar surface (NRC 2007; Flahaut et al. 2012; Crawford et al. 2012). These include determining whether there was, or was not, a catastrophic spike in the impact rate between 3.8 and 4.0 Gyr ago (i.e. a so-called Late Heavy Bombardment, with implications for both conditions on the early Earth and outer planet orbital dynamics; e.g. Levison et al. 2001; Chapman et al. 2007); the inner Solar System cratering rate (and thus planetary surface age determination) in the range 1-3 Gyr ago that was not well sampled by Apollo (Fig. 4); the record of ancient solar wind and galactic cosmic rays (with their record of solar evolution and the changing galactic environment of the Solar System; see Crawford et al 2010 and references therein); and the sampling of 'exotic' lunar lithologies not represented in the Apollo sample collection, including samples originating from the deep lunar interior. It has also become clear that the lunar surface, especially the farside, would be an excellent location for low-frequency radio astronomy (e.g. Jester and Falcke, 2009), and various astrobiological and life sciences investigations (e.g. Cockell 2010).

Some of these future studies could undoubtedly be performed with targeted robotic landers dispatched to key localities, such as ESA's proposed Lunar Lander (Carpenter et al. 2012) and the proposed MoonRise sample return mission (Jolliff et al 2010). However, if Apollo taught us anything regarding planetary exploration it is that, expensive though human exploration certainly is, the sheer *efficiency* of having people on site exploring planetary surfaces sufficiently transcends what can be accomplished robotically that science is a net beneficiary. Apollo also taught us that, in addition to advancing science, large-scale human space missions are effective at driving technological innovation, at inspiring young people to become interested in science and exploration (the current author among them), and in drawing people together through a sense of our common humanity in a cosmic setting.

Therefore, as we pass the 40th anniversary of the last human expedition to the Moon, and mark the passing of the first person ever to have set foot upon its surface, for both scientific and societal reasons now is an appropriate time to start serious planning for a return. However, unlike the Cold War competition that drove Apollo, a human return to the Moon in the coming decades would ideally be part of a sustained, international, programme of Solar System exploration such as that envisaged in the recently formulated Global Exploration Roadmap (ISECG 2011).

**Acknowledgements**

I thank Dr Katherine Joy for comments on an earlier version of the manuscript which have improved it, Dr Claire Cousins for chasing up statistics related to our Icelandic fieldwork, and the Apollo Sample Curator, Dr Ryan Zeigler, for information on CAPTEMs recent sample allocations.